\documentclass[%
 reprint,
 amsmath,amssymb,
 aps,
]{revtex4-2}
\usepackage[english]{babel}
\usepackage{graphicx}
\usepackage{dcolumn}
\usepackage{bm}
\usepackage{amssymb}
\usepackage{amsfonts}
\usepackage[sort&compress]{natbib}
\usepackage{color}

\usepackage{xcolor}

\begin{document}
\preprint{APS/123-QED}
\title{Inertia onset in disordered porous media flow}
\author{Damian Śnieżek}
\email{damian.sniezek@uwr.edu.pl}
\author{Sahrish B. Naqvi}%
 \email{sahrishbatool.naqvi@uwr.edu.pl}

\author{Maciej Matyka}%
 \email{maciej.matyka@uwr.edu.pl}
\affiliation{%
 Institute of Theoretical Physics, University of Wroc{\l}aw, pl. M. Borna 9, 50-204 Wroc{\l}aw, Poland
}%

\date{\today}

\begin{abstract}
We investigate the very onset of the inertial regime in the fluid flow at the pore level in a three-dimensional, disordered, highly porous media. 
We analyze the flow structure in a wide range of Reynolds numbers starting from 0.01 up to 100. We focus on qualitative and quantitative changes that appear with increasing Reynolds number. To do that, we investigate the weakening of the channeling effect, defined as the existence of preferred flow paths in a system. We compute tortuosity, spatial kinetic energy localization, and the pore-space volume fraction containing negative streamwise velocity to assess accompanying changes quantitatively.
Our results of tortuosity and participation number derivatives show that the very onset of inertia is apparent for Reynolds number Re $\sim 0.1$, an order of magnitude lower than indicated by analyzing relations of friction factor with the Reynolds number. Moreover, we show that the vortex structures appear at Reynolds number two orders of magnitude higher than the onset of inertia.\end{abstract}
\maketitle

\section{Introduction} One of the most striking phenomena in transport through porous media is the formation of flow channels, determined not only by the pore geometry but by the transport phenomenon itself \cite{Hyman2020}. Similar phenomena occurs in two-phase fluid flow as shown experimentally in \cite{tallakstad2009steadyprl, tallakstad2009steadypre}. Several factors influence fluid behavior in porous media, such as pore size distribution, porosity, permeability, and the complexity of fluid pathways, often referred to as tortuosity \cite{matyka2008tortuosity}. Its understanding is fundamental in hydrology, petroleum engineering, environmental science, and chemical engineering. Applications of porous media span across various fields, including microfluidic devices \cite{cao2019application}, energy production and development \cite{banerjee2021developments}, and the design of filters \cite{zhang2024optimal} and chemical reactors \cite{wang2014thermal}, just to name a few. Usually, there are 
two empirical laws used to describe fluid dynamics in porous media. 
In creeping, laminar regime, \cite{neuman1977theoretical, whitaker1986flow} Darcy's law $\mathbf{u}=-\frac{\mathcal{K}}{\mu L} \nabla p$ is used, where $\mathbf{u}$ is the velocity vector, $\mu$ the dynamic viscosity of the fluid, $\mathcal{K}$ the permeability of the domain, $L$ the streamwise length of porous sample, and $p$ the pressure.
However, the deviation from linear Darcy's law are observed for lower \cite{roy2021role, roy2019effective} and higher velocities. In the latter, the nonlinear relationship between the pressure drop and the filtration velocity (or flowrate) is described by introducing the Forchheimer term, which is the second order correction to Darcy law: $\mathbf{u}=-\frac{\mathcal{K}}{\mu L} \nabla p-\beta \mathcal{K}|\mathbf{u}| \mathbf{u}$, \cite{basak1977non, whitaker1996forchheimer, zeng2006criterion, Arbabi2024}, where $\beta$ is a model parameter scaling the quadratic velocity term.
The impact of large-scale permeability heterogeneity on high-velocity flows in porous media was analyzed in \cite{fourar2005inertia}. There, an effective inertial coefficient, which was determined from numerical simulations, varied with Reynolds number, variance, and mean permeability ratio differently across serial-layered, parallel-layered, and correlated media.

The boundaries between Darcy and inertial flow regimes in two-dimensional porous media were experimentally observed for the first time in the rod porous medium model~\cite{dybbs1984new}. It was shown that the transition between linear and nonlinear regimes occurs for $1<Re<10$. Similar range of Reynolds Number $0.83 < Re < 9.2$ was reported in~\cite{wahyudi2002darcy}, where fluid flow through five different, three dimensional, sand samples was analyzed experimentally. Forchheimer's flow through four different permeable stones' samples was investigated and the influence of porosity, mean pore size and pore size distribution on transition to non-Darcy flow was discussed in ~\cite{li2022experimental}. In two-dimensional, disordered porous media, $Re=0.37$ was indicated~\cite{andrade1999inertial}. A study utilizing network modeling to analyze the onset of non-Darcy flow in porous media highlighted the significant influence of pore geometry on the critical Reynolds number and the onset of inertial flow~\cite{veyskarami2018new}. The influence of high fluid velocities on fluid flow and tracer transport in heterogeneous porous media was investigated in~\cite{nissan2018inertial}. An experimental study of two-phase flow through a porous sandstone sample reports the existence of three flow regimes. The authors detect the onset of inertia between the capillary-dominated and intermittent flow regimes~\cite{gao2020pore}. Structural complexity in the media was shown to intensify inertial effects and lead to a more homogenized flow~\cite{zolotukhin2022analysis}. In consequence, predictable chemical transport patterns were observed.

Our aim is to take an insight into the physical mechanisms that govern the transition of the pore-scale flow to a nonlinear regime.
In this work, we study the flow through disordered porous media. We perform fully converged numerical simulations of Navier-Stokes equations in pore scale. In particular, we analyze flow structure, tortuosity, and spatial distribution of kinetic energy at the very onset of the inertial regime. 

\section{Model and Method} 

Our porous media samples consist of spherical non-overlapping obstacles with a diameter of $D=1$ randomly distributed in a system with porosity $\varphi=0.9$ of size $L=(x\times y\times z)=(16D\times16D\times16D)$. 
The sample was placed in the $(66D\times16D\times16D)$ empty channel at $x=20D$ in order to minimize effect of inlet and outlet boundary conditions on the pore-scale flow.
We generated eight independent random samples. The steady, laminar, incompressible Navier-Stokes equations govern the flow in pore-space:
\begin{equation}
    \nabla \cdot \mathbf{u}=0
    \label{eq:1}
\end{equation}
\begin{equation}
    \rho \left(\vec{u} \cdot \nabla\right) \vec{u}=-\nabla p+ \mu \nabla^2 \vec{u},
    \label{eq:2}
\end{equation}
where $\rho$ is the density of fluid, $\mu$ is kinematic viscosity of the fluid, $\vec{u}=(u_1,\,u_2,\, u_3)$ is the pore-scale velocity field, and $p$ is the pressure. We set a constant, uniform velocity  $u_{in}$ at the inlet and a zero pressure gradient at the outlet. All other boundaries, including the obstacles, are treated as impremeable solids with no-slip boundary condition. We varied inlet velocity $u_{in}$, to cover the range of Reynolds numbers $0.01\leq Re \leq 100$, where $Re=\rho u_{in} D/\mu$. 
We solved Eqs.~\eqref{eq:1} and ~\eqref{eq:2} in pore-space using a standard finite volume method. For this we employed a finite volume-based semi-implicit method for pressure-linked equations (SIMPLE algorithm) within the OpenFOAM framework \cite{jasak2007openfoam}. Our criterion for determining steady-state convergence in the simulations was set to $\epsilon \leq 10^{-6}$, where $\epsilon$ was the normalized difference between velocity magnitudes in successive iterations. All simulations were executed in parallel mode using 8-core processors within the OpenMPI library. Domain discretization was accomplished using the standard meshing tools blockMesh and snappyHexMesh of OpenFOAM.
To ensure numerical reproducibility, we adopted the standard Richardson extrapolation technique \cite{roache1997quantification,cadafalch2002verification}.

\section{Results and Discussions}
\begin{figure}
    \centering
\includegraphics[width=1\linewidth]{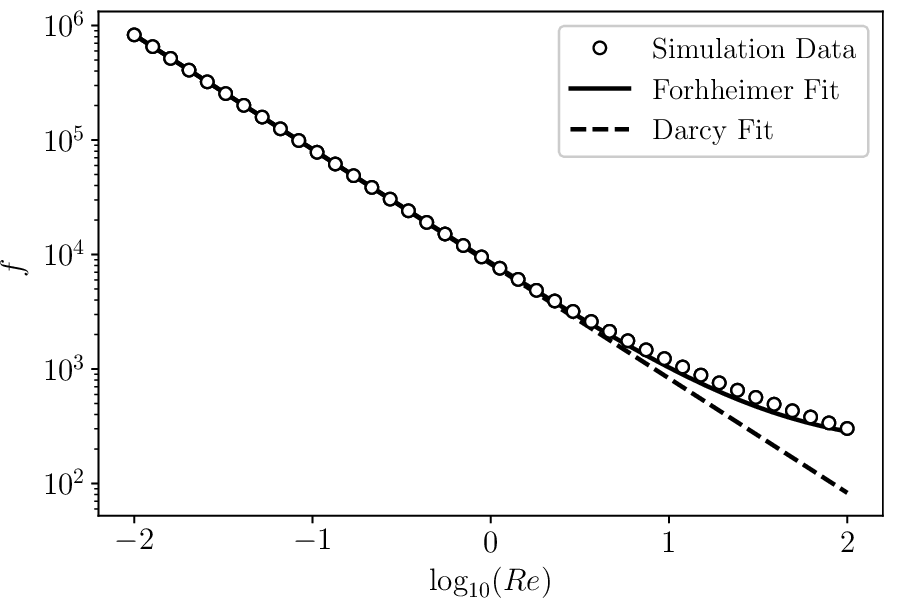}
\caption{\label{fig:ff} Generalized friction factor $f$ in relation to the Reynolds number. The solid and dashed lines represent the best fit to the Forchheimer and Darcy equations, respectively. The error bars
are smaller than the Simulation Data symbols.}
\end{figure}

First, we identify the flow deviation between Darcy and the inertial regime by calculating the friction factor ($f \equiv -\Delta p / D \beta \rho u^2$) \cite{andrade1999inertial} for various Reynolds numbers, shown in Fig.~\ref{fig:ff}. 
The results are presented as averages from eight independent samples along with the standard error of the mean.
This result shows the regime where Darcy's law correctly describes the system. Here, the transition regime appears between  $1\leq Re \leq 10$, which is in agreement with experimental results~\cite{wahyudi2002darcy}.

Next, we compute tortuosity to quantify the elongation of fluid pathways ~\cite{{green1951fluid},{thauvin1998network},{cooper1999non},{coles1998non}} and their change due to the appearance of inertia. We use the method presented in ~\cite{matyka2008tortuosity}, where it is defined as:

\begin{equation}
   \tau=\frac{\langle|\vec{u}|\rangle}{\langle u_1\rangle},
   \label{eq:T}
\end{equation}
where $\langle|\vec{u}|\rangle$ represents the average magnitude of the fluid velocity and $\langle u_1 \rangle$ represents the average streamwise velocity component $u_1$ in the direction of the flow across the porous sample.
\begin{figure}
    \centering
     \includegraphics[width=1\linewidth]{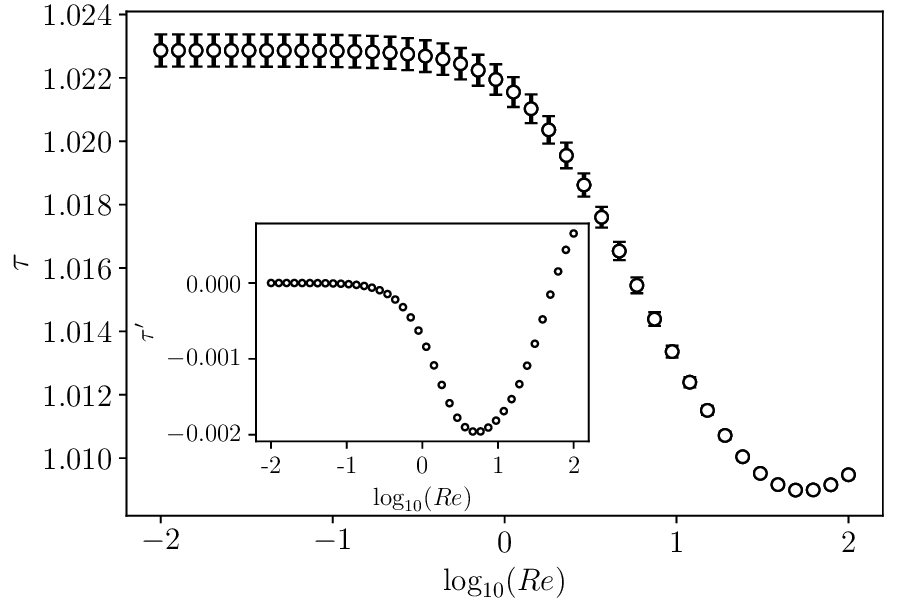}
    \caption{Tortuosity ($\tau$) and its derivative (\mbox{$\tau^{\prime}=d\tau/d\mathrm{log}_{10}Re$}) (inset).}
    \label{fig:T}
\end{figure}
In Fig.~\ref{fig:T}, we plot tortuosity $\tau$ versus $Re$. 
Tortuosity remains constant in the Darcy regime.
It starts decreasing approximately at $Re=1$, which agrees with \cite{sivanesapillai2014transition}. Initially, when the viscous forces are dominating, flow closely follows the pore structure (tortuosity is constant).
However, as $Re$ increases and the flow becomes more inertial, the shortening of path lines is visible as tortuosity decreases.
The value of $\tau$ keeps decreasing till it reaches the minimum value at $Re\approx62$. After this point, a slight increase is observed, possibly due to the appearance of vortices. Moreover, the decrease of the standard error with increasing $Re$ is noticeable. This suggests a weaker influence of geometry details on the flow structure in higher Reynolds numbers regime. 
We further show the derivative $\tau^{\prime}$ (inset in Fig.~\ref{fig:T}) to identify the Reynolds number at which tortuosity starts changing (Fig.~\ref{fig:T}), which is at $Re\approx0.1$. 

\begin{figure}
    \centering
    \includegraphics[width=1\linewidth]{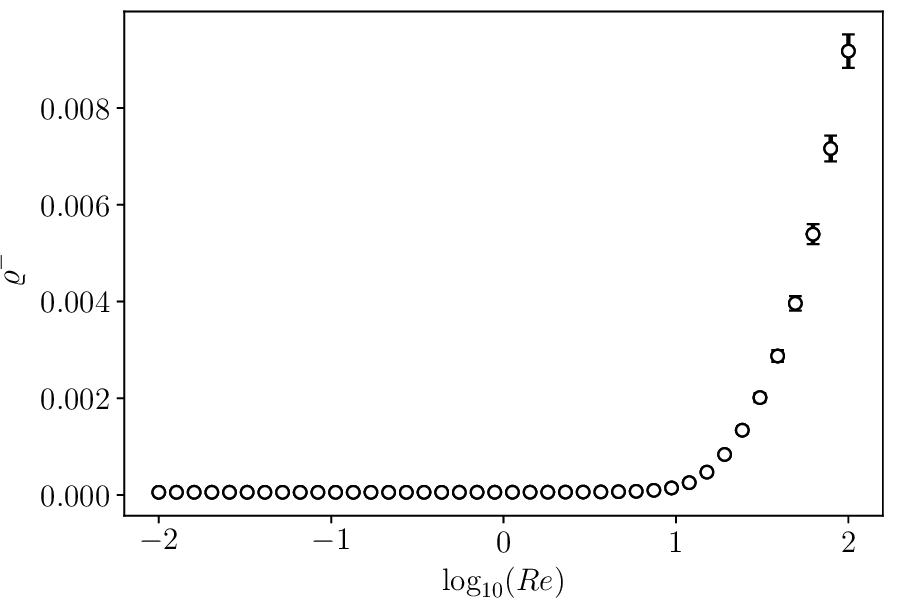}
    \caption{Volume fraction of the pore space with negative streamwise velocity.}
    \label{fig:vdf}
\end{figure}

To understand the physical mechanism of tortuosity decrease at the onset of the inertial regime, we compute the volume fraction of the pore space with negative streamwise velocity $u_1$ as:
\begin{align}
 \varrho^{-} &=\frac{1}{V_{\text{fluid}}} \int_{V_{\text{fluid}}} f\left(u_1(\vec{r})\right)d^3 r \nonumber \\
&\approx \frac{1}{V_{\text{fluid}}} \sum_i^n f\left(u_1^i\right) V_i \text{,} \label{eq:3}
\end{align}
where $i$ denotes mesh's cell index, $n$ is the total number of cells, $V_i$ is the $i$-th cell volume. The function $f$ is defined as: 
\begin{equation}    
f\left(u_1\right)=\left\{\begin{array}{cc}
1, & u_1<0 \\
0, & \text { otherwise }.
\end{array}\right.
\end{equation}
The results of $\varrho^{-}$ are illustrated in Fig.~\ref{fig:vdf}. Interestingly, the Reynolds number at which the backward flow emerges ($Re\approx10$) is not compliant with Reynolds number at which inertia appears, indicated in Fig.~\ref{fig:ff} and Fig.~\ref{fig:T}. To understand this, we calculate the spatial distribution of kinetic energy within the porous samples, a statistical measure of the channeling effect \cite{andrade1999inertial}.
To quantify this phenomenon, we compute the energy localization defined as participation number $\pi$ \cite{andrade1999inertial}:
    \begin{equation}
    	\label{eq:pi_final}
    	\pi = \left(n\sum_{i=1}^{n}q_i^2\right)^{-1}, 
    \end{equation}
\begin{figure}[h!]
    \centering
    \includegraphics[width=1\linewidth]{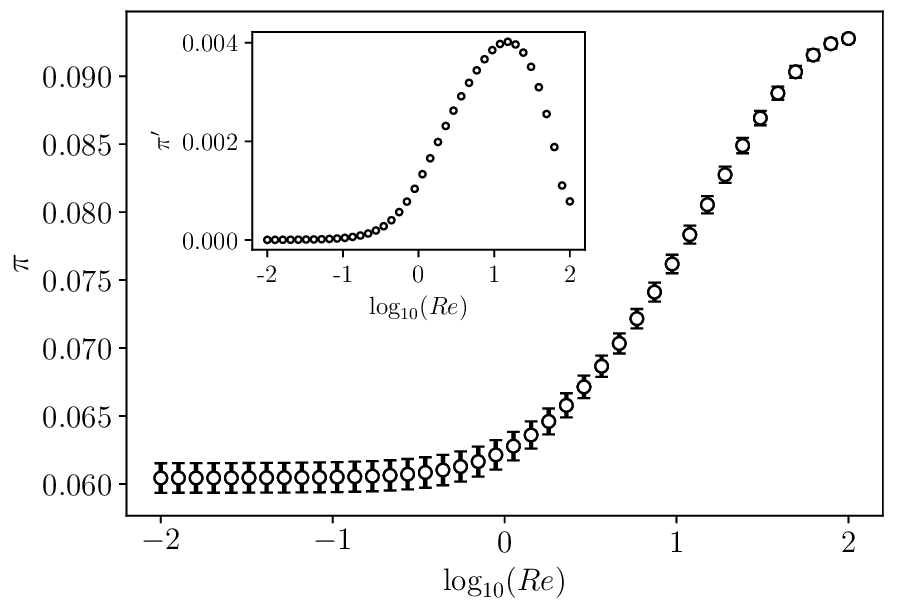}
    \caption{Participation number ($\pi$) and its derivative (\mbox{$\pi^{\prime}=d\pi/d\mathrm{log}_{10}Re$}) (inset).}
    \label{fig:PI}
\end{figure}
where $q_i=e_i/\sum_{j}^{n} e_j$ is the fraction of the total pore space kinetic energy contained in the $i$-th mesh cell.
Results of energy distribution are given in Fig.~\ref{fig:PI}.
The plot shows a rise of $\pi$ with the Reynolds number.
High value of $\pi$ indicates a more homogeneous energy distribution, which
represents the weakening of the channeling effect. 
\begin{figure}[h!]
    \centering
    \includegraphics[width=0.9\linewidth]{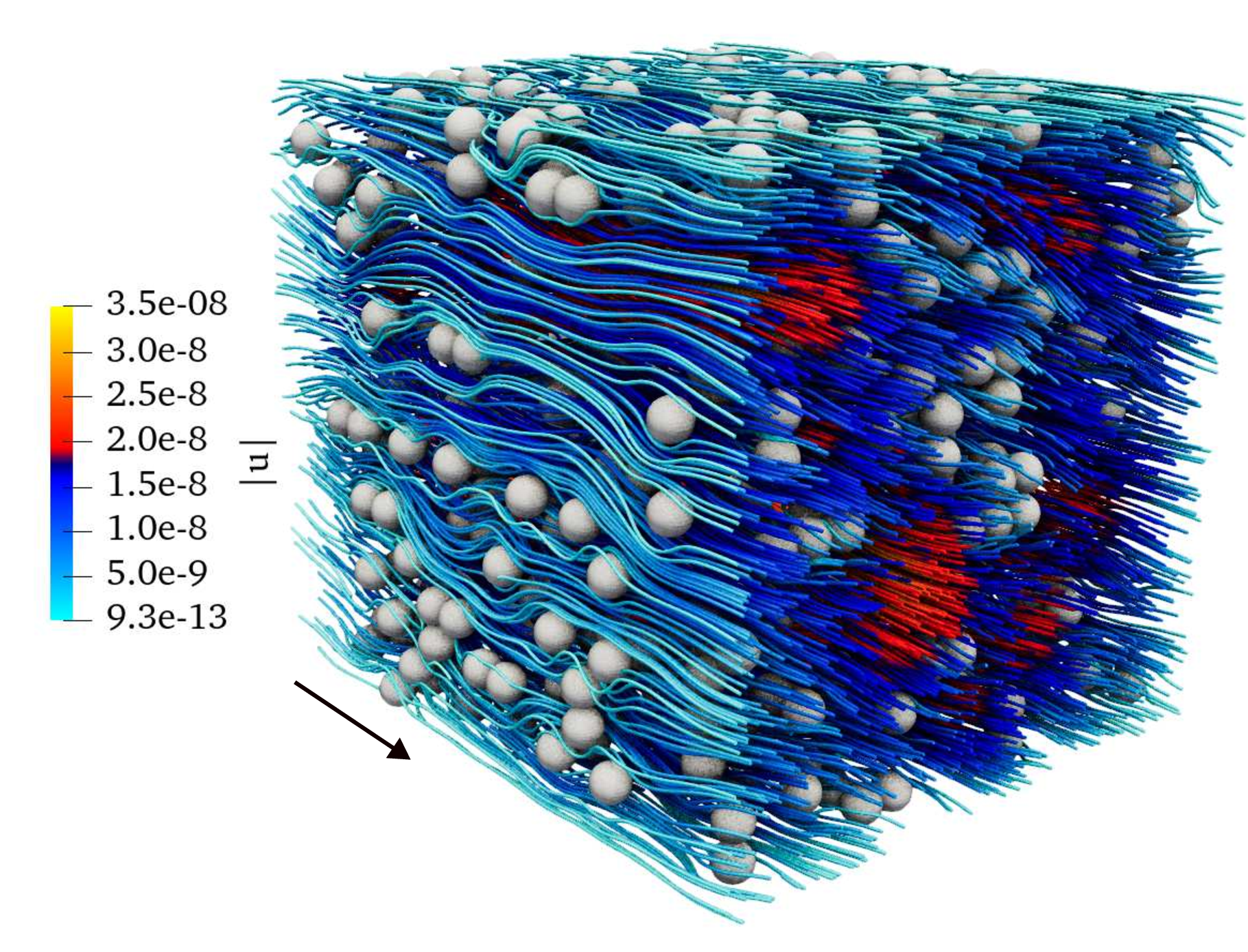}\\
    \includegraphics[width=0.9\linewidth]{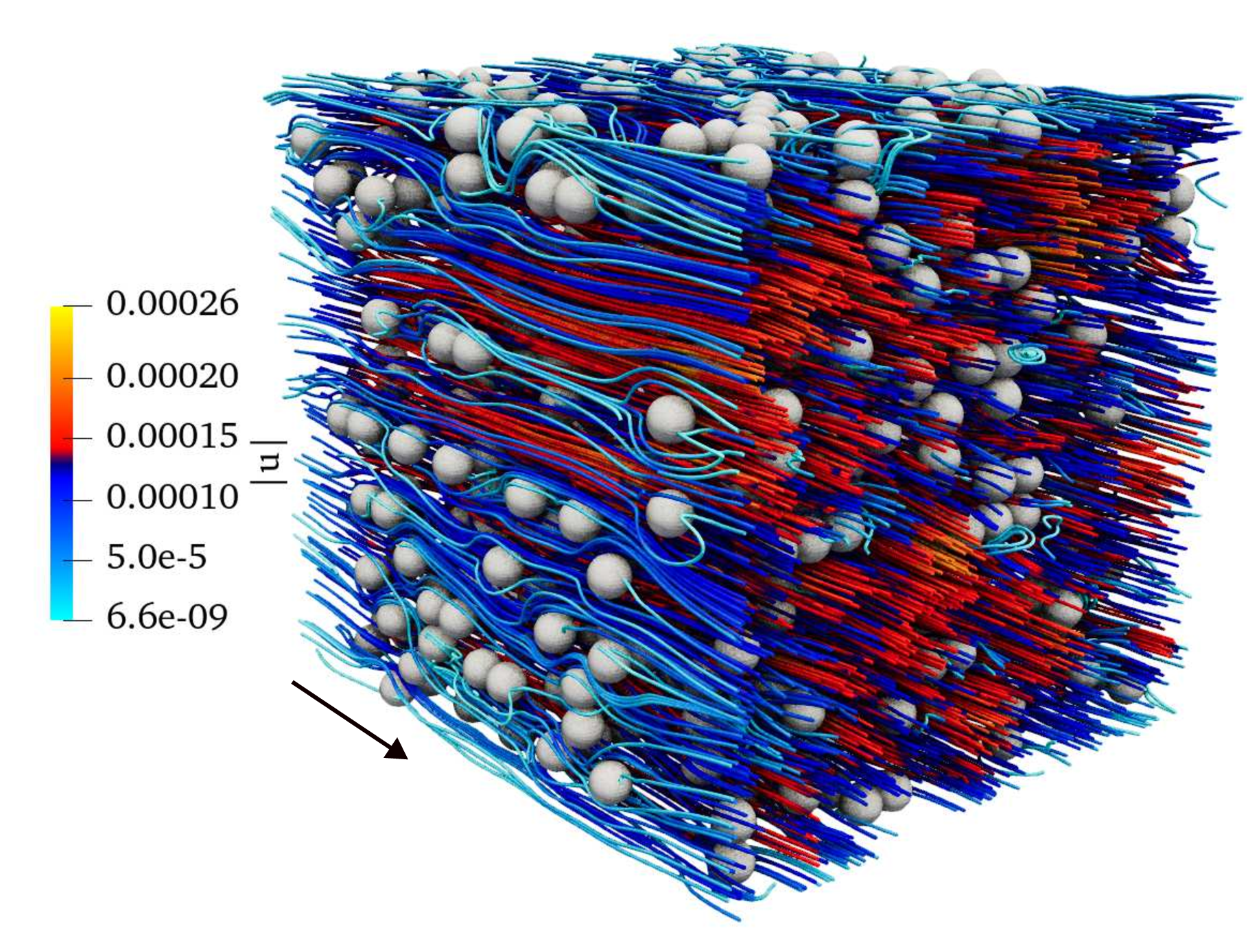}\\
    \caption{Flow streamlines (colored by velocity magnitude or color intensity in gray-scale) for two Reynolds numbers: $Re=0.01$ (top) and $Re=100$ (bottom). The obstacles are shown in white color. The arrows indicate the flow direction.
    }
    \label{fig:stream}
\end{figure}
This effect is clearly visible when one compares the streamlines in low and high Re regimes (see Fig.~\ref{fig:stream}).
The strong channeling effect at a low Reynolds number ($Re=0.01$) is indicated by the fact that only small part of pore space volume give a significant contribution to the transport. 
With increasing $Re$, as the kinetic energy is more dispersed in pore space, weakening of the channeling effect is observed. 
This is visible as the change in colour of streamlines (decreased intensity in grayscale) in Fig.~\ref{fig:stream}. A more dispersed flow channels are clearly visible at higher Reynolds number ($Re=100$). This shift from preferential channeling to more dispersed flow patterns 
as $Re$ increases is due to the 
inertial forces surpassing viscous forces.
It causes the fluid flow to spread out and fill more of the available pore space \cite{andrade1999inertial}. This agrees with our previous observation in $\pi$ increase (see Fig.~\ref{fig:PI}). We also observe the presence of backward flow at $Re=100$, indicated by elongated, highly tortuous streamlines in low-velocity regions which agrees with our results of negative velocity appearance at increasing $Re$ given in Fig.~\ref{fig:vdf}. Here, fluid particles are transported along paths that diverge from the previously geometry guided channels. The occurrence of backward flow, and thus vortices, suggests complex flow interactions and is a precursor to turbulent effects that may appear if the Reynolds number increases further.

\section{Conclusions}
In that study, we investigated what physical phenomena drive the transition from Darcy to the inertial regime in the flow through disordered porous media.

By examining Reynolds numbers from $0.01$ to $100$, we identify significant qualitative and quantitative changes in the flow structure as inertia becomes significant. Our findings indicate that the onset of inertia, characterized by a weakening of the channeling effect and the appearance of vortices, occurs at a Reynolds number ($Re$) as low as 0.1. This is an order of magnitude lower than the traditionally accepted value derived from the friction factor-Reynolds number relationship. This early onset of inertial effects is marked by changes in tortuosity and the spatial distribution of kinetic energy, with tortuosity beginning to decrease at $Re \approx 1$. The study also reveals that the volume fraction of the pore space containing negative streamwise velocity emerges at $Re \approx 10$, indicating the formation of backward flows and vortices, which become more pronounced at higher Reynolds numbers. These phenomena are further confirmed by the increase in the participation number, suggesting a more homogeneous energy distribution and a reduction in the channeling effect. The visualization of streamlines supports this conclusion, showing a transition from highly channeled to more dispersed flow patterns as $Re$ increases. The emergence of backward flow and the resulting vortices at higher $Re$ indicate complex flow interactions, hinting at the precursor to turbulence in porous media. Our study highlights the nuanced and early onset of inertial effects in disordered porous media, providing a deeper understanding of the transition from Darcy to non-Darcy flow regimes. Our findings underscore the importance of considering pore-scale flow dynamics in studying fluid flow in the inertial regime in porous systems.

\section*{Acknowledgement}

Funded by National Science Centre, Poland under the OPUS call in the Weave programme 2021/43/I/ST3/00228.
This research was funded in whole or in part by National Science Centre (2021/43/I/ST3/00228). For the purpose of Open Access,
the author has applied a CC-BY public copyright licence to any Author Accepted Manuscript (AAM) version arising from this submission.

The authors are thankful for Dawid Strzelczyk's support, his in-depth discussions, and insights into this work.
We are also grateful to Jos\`e S. Andrade Jr. for insightful discussion and comments during the early stages of the work.

%

\end{document}